\documentclass[10pt,onecolumn,letterpaper]{IEEEtran}
\usepackage{enumerate}
\usepackage{subfigure}
\usepackage{multirow}
\usepackage{url}
\usepackage{array}
\usepackage{amsmath}
\usepackage{amssymb}
\usepackage{multicol}
\usepackage{soul}
\usepackage{float}
\usepackage{multirow}
\usepackage{multicol}
\usepackage{extarrows}
\usepackage{mathrsfs}
\usepackage{bm}
\usepackage{color}
\usepackage{booktabs}
\usepackage[sort]{cite}

\begin{document}

\title{CNN Based Adversarial Embedding with
Minimum Alteration for Image Steganography}

\author{Weixuan Tang,
        Bin Li*,
        Shunquan Tan,
        Mauro Barni,
       and Jiwu Huang

\thanks{W. Tang, B. Li, S. Tan, and J. Huang are with Guangdong Key Laboratory of Intelligent Information Processing and Shenzhen Key Laboratory of Media Security, Shenzhen University, Shenzhen 518060, China (email: tweix@mail2.sysu.edu.cn; libin@szu.edu.cn;  tansq@szu.edu.cn; jwhuang@szu.edu.cn).}
\thanks{M. Barni is with Department of Information Engineering and Mathematics, University of Siena, Siena 53100, Italy (email: barni@dii.unisi.it).}
\thanks{W. Tang is also with Sun Yat-sen University, Guangdong 510006, China.}\thanks{*B. Li is the correspondence author.}
}


\maketitle

\begin{abstract}
Historically, steganographic schemes were designed in a way to preserve image statistics or steganalytic features. Since most of the state-of-the-art steganalytic methods employ a machine learning (ML) based classifier, it is reasonable to consider countering steganalysis by trying to fool the ML classifiers.
However, simply applying perturbations on stego images as adversarial examples may lead to the failure of data extraction and introduce unexpected artefacts detectable by other classifiers.
In this paper, we present a steganographic scheme with a novel operation called adversarial embedding,
which achieves the goal of hiding a stego message while at the same time fooling a convolutional neural network (CNN) based steganalyzer.
The proposed method works under the conventional framework of distortion minimization. Adversarial embedding is achieved by adjusting the costs of image element modifications according to the gradients backpropagated from the CNN classifier targeted by the attack.
Therefore, modification direction has a higher probability to be the same as the sign of the gradient.
In this way, the so called adversarial stego images are generated.
Experiments demonstrate that the proposed steganographic scheme is secure against the targeted adversary-unaware steganalyzer. In addition, it deteriorates the performance of other adversary-aware steganalyzers opening the way to a new class of modern steganographic schemes capable to overcome powerful CNN-based steganalysis.
\end{abstract}


\IEEEpeerreviewmaketitle

\section{Introduction}
\label{sec:intro}
Image steganography is the art and science of concealing covert information within images. It is usually achieved by modifying image elements, such as pixels or DCT coefficients. On the other side of the game, steganalysis aims to reveal the presence of secret information by detecting whether there are abnormal artefacts left by data embedding.

The developing history of  steganography and steganalysis is rich of interesting stories,
as they compete with each other and they benefit and evolve from the competition\cite{Li2011}.
The earliest steganographic method was implemented by substituting the least significant bits of image elements with message bits. The stego artefacts introduced by this method can be  effectively detected by Chi-squared attack  \cite{Chi_squared}, or steganalytic features based on \textit{first-order statistics} \cite{WS}.
In this initial phase of the competition, statistical hypothesis testing or a simple linear classifier such as FLD (Fisher Linear Discriminant) could serve the need of steganalysis.

The first-order statistics can be restored after data embedding, as done in \cite{OutGuess}.
The abnormal artefacts in the first-order statistics can also be avoided as in \cite{LSBMatching, luo2010edge}.
As a consequence, more powerful steganalytic features based on the \textit{second-order statistics} \cite{SPAM, ChunHua} were proposed.
In  this period, advanced machine learning (ML) tools, such as SVM (Support Vector Machine), were  operated on high-dimensional features (where the dimension is typically several hundreds).
These methods were very effective in detecting steganographic schemes even if the first-order statistics were preserved.

Modern steganographic schemes are designed under the framework of \textit{distortion minimization} \cite{simulator}.
The embedding cost of changing each image element is specified by a cost function, and a coding scheme is employed to convey information by minimizing the distortion, which is computed as the total cost of modified elements.
For example,
the schemes in \cite{HUGO, WOW, UNIWARD, HILL, UED, MP}
are well-known for their elegant cost functions.
As a counter measure, state-of-the-art steganalytic methods adopt
\textit{higher-order statistics} with much higher dimensional features (where the dimension is typically thousands or even more than ten thousands), such as in \cite{SRM, TLBP, JRM, DCTR, GFR}. More sophisticated ML methods, such as the ensemble classifier \cite{Ensemble}, have also been employed.

Steganalytic methods based on deep learning \cite{Tan,Qian,Xu,Xu2,XuJPEG,zeng} have rapidly gained an increasing attention in recent years.
Without the need of designing hand-crafted features,
deep convolutional neural networks (CNN) show a promising way in automatic feature extraction and classification for steganalysis.
Incorporated some domain knowledge into the network design,
such as using high-pass
filters for pre-processing,
outstanding performance can be obtained.

The high-dimensional hand-crafted or deep-learned features with the powerful supervised ML schemes present a great challenge to steganography.
A promising strategy for the steganographer is to use side information which is not available to the steganalyst, such as using the camera sensor noise during message embedding \cite{JEPGNatural} and the compression noise during JPEG compression \cite{UNIWARD}.
However, the side information is not always available for all kinds of cover images, especially for those already compressed in JPEG format.
As a consequence, better steganographic schemes suitable for more general conditions are needed.

As the dimension of steganalytic features increases, it is difficult for steganograhpy to preserve all statistical features during data embedding.
This motivates us to find a better way to resist steganalysis by countering the ML based classifier.
Recent studies \cite{fool, property} have shown that ML systems are vulnerable to intentional {adversarial operations}.
For example, Chen \textit{et al.} \cite{forensics} have shown that the performance of an image forensics detector with a SVM classifier can be greatly degraded by a rather simple gradient based attack.
There is  also some research evidence indicating that classifier based on deep learning can be easily fooled by
\textit{adversarial examples} \cite{AD,deepFool,resistant},
which are formed by applying small but intentional
perturbations to inputs in order to make the classification model yield erroneous outputs.
However, applying adversarial perturbations as in \cite{AD} on stego
images may lead to data
extraction failures. The perturbations may also introduce unexpected artefacts detectable by other
classifiers.

The progress in adversarial signal processing \cite{Barni2013a} inspired us to design a steganographic scheme that is resistant against ML based steganalyzers.
In this paper, we propose a scheme called
AMA (\underline{A}dversarial embedding with \underline{M}inimum \underline{A}lternation).
Targeted to counter Xu-CNN JPEG steganalyzer \cite{XuJPEG},
we generate a new kind of stego images via \textit{adversarial embedding},
an operation that takes into account both the embedding of the stego message and the necessity to fool the targeted steganalyzer.
AMA is implemented under the framework of distortion minimization,
and based on a baseline steganographic scheme adopting a conventional embedding mechanism.
Specifically, AMA adapts the cost assignment process by asymmetrically adjusting a portion of embedding costs according to the gradients backpropagated from the deep learning steganalyzer.
In order to avoid unnecessary extra modifications, the amount of image elements with adjustable costs is kept to a minimum.
Experimental results show that the \textit{adversarial stego images} generated by AMA with adversarial embedding can successfully fool the targeted deep learning steganalyzer,
which was trained with several hundreds of thousands of training images.
More interestingly, although the adversarial stego images have a higher rate of modifications, they are less detectable by other advanced hand-crafted feature based steganalyzers than stego images generated by the baseline steganographic scheme.
These results seem to suggest that, guided by necessity to fool the data-driven deep learning steganalyzer, adversarial embedding implicitly preserves the image statistics to some extent.

The main contributions of our work are as follows:
\begin{enumerate}
  \item A new strategy to fool the ML classifiers, which is not based on the attempt to preserve a specific image statistical model, is proposed. We believe this is a promising way to counter steganalysis.
  \item A practical steganographic scheme called AMA with adversarial embedding operation is proposed. As opposed to conventional approaches used to cerate adversarial examples in other machine learning domain, adversarial stego images generated by the AMA scheme are capable of carrying secret information.
  \item Based on the knowledge available to the steganographer and the steganalyst, different adversarial models are considered, wherein the proposed scheme can achieve state-of-the-art security performance.
\end{enumerate}

The rest of the paper is organized as follows.
In Section \ref{sec:problem}, we formulate the problem of steganography and steganalysis.
We point out the foundation of the proposed steganographic scheme,
and differentiate two kinds of adversarial scenarios.
We present the idea as well as a practical implementation of the proposed AMA steganographic scheme in Section \ref{sec:amr}.
Extensive experiments are performed and the results are reported in Section \ref{sec:exp} to demonstrate the performance of the AMA scheme
under different adversarial conditions when compared to the baseline steganographic method.
Conclusions are presented in Section \ref{sec:con}.

\section{Problem Formulation}
\label{sec:problem}
In this article, capital letters in bold are used to represent matrices.
The corresponding lowercase letters are used for matrix elements.
The flourish letters are used for sets.
Specifically, cover and stego images are respectively denoted as $\mathbf C = (c_{i,j})^{H \times W }$and $\mathbf S =(s_{i,j})^{H \times W }$, where $H$ and $W$ are the height and width of the image. Their corresponding image sets are denoted as $\mathcal{C}$ and $\mathcal{S}$.
In order to differentiate the proposed \textit{adversarial stego images} from conventional stego images, we use $\mathbf Z = (z_{i,j})^{H \times W } \in \mathcal{Z}$. Note that $\mathbf Z$ is a special type of $\mathbf S$.

\subsection{Practical Evaluation Metrics for Steganographic Security}
\label{subsecII:evaluation}
The fundamental requirement of steganalysis is to differentiate stego images from cover images.
When analyzing an image $\mathbf X$,
the steganalyst must decide between the following two hypotheses:
\begin{equation}
\begin{aligned}
H_{0}: \ & \mathbf X \text{ is a cover image.}\\
H_{1}: \ & \mathbf X \text{ is a stego image.}\\
\end{aligned}
\end{equation}
To accomplish this task in a supervised ML setting, the steganalyzer would train a classifier $\phi_{\mathcal{C,S}}$ with binary output using training data from $\mathcal{C}$ and $\mathcal{S}$, and obtain the decision criterion as follows:
\begin{equation}
\begin{aligned}
\begin{cases}
\mathbf X \text{ is a cover image,}  &\text{if $\phi_{\mathcal{C,S}}(\mathbf X)=0$},\\
\mathbf X \text{ is a stego image,}  &\text{if $\phi_{\mathcal{C,S}}(\mathbf X)=1$}. \\
\end{cases}
\end{aligned}
\end{equation}
The trained classifier is called \textit{steganalyzer}.
Two types of errors can occur. \textit{Type I error} is the missed detection where stego images are misclassified, and \textit{Type II error} is the false alarm where cover images are misclassified.
Their corresponding error probabilities are defined as:
\begin{equation}\label{eq:md}
P_{md}^{\text{$\phi_{\mathcal{C,S}}$}} = \Pr \{\text{$\phi_{\mathcal{C,S}}(\mathbf S)=0$} \},
\end{equation}
and
\begin{equation}\label{eq:fa}
P_{fa}^{\text{$\phi_{\mathcal{C,S}}$}} = \Pr \{\text{$\phi_{\mathcal{C,S}}(\mathbf C)=1$} \}.
\end{equation}
Under equal Bayesian prior for cover and stego,
the total error rate is
\begin{equation}\label{eq:te}
\begin{aligned}
P_{e}^{\text{$\phi_{\mathcal{C,S}}$}} = \frac{P_{md}^{\text{$\phi_{\mathcal{C,S}}$}}+P_{fa}^{\text{$\phi_{\mathcal{C,S}}$}}}{2}.
\end{aligned}
\end{equation}
The goal of the steganalyst is to minimize $P_{e}^{\text{$\phi_{\mathcal{C,S}}$}}$, while
that of the steganographer is to maximize it.

\subsection{Steganographer's Knowledge about Steganalyzer}
\label{subsecII:steganography}
The steganographer may have different levels of knowledge about $\phi_{\mathcal{C,S}}$, such as the classification scheme and the training data.
In this paper, we will not discuss what is the best strategy the steganographer should take according to the accessibility of these knowledge.
Instead, we assume the gradients of the loss function with respect to the input, which are
backpropagated from a ML based steganalyzer $\phi_{\mathcal{C,S}}$, are accessible to the steganographer.
This is the foundation of the proposed steganographic scheme.
In Section \ref{sec:amr},
we will propose a scheme to fool such a steganalyzer with \textit{adversarial stego images}.
We will also investigate in the experimental part
how the adversarial stego images perform under other advanced steganalyzers
(\textit{e.g.}, $\phi_{\mathcal{C,S}}^{\prime}$)
where the knowledge of these steganalyzers is unavailable.

\subsection{Steganalyst's Knowledge about Adversarial Stego Images}
\label{subsecII:steganalysis}
If a steganalyst is unaware of any adversarial operation,
he is called \textit{adversary-unaware steganalyst}.
Otherwise, he is called \textit{adversary-aware steganalyst}.
The best reaction of an adversary-aware steganalyst is to re-train the classifier with adversarial stego samples
to obtain a new steganalyzer $\phi_{\mathcal{C,Z}}$,
or use other advanced steganalyzers (\textit{e.g.,} $\phi_{\mathcal{C,Z}}^{\prime}$) unknown to the steganographer.
This may present two most challenging cases for a steganographer and we will discuss these scenarios in the experiments.

\section{The Proposed AMA Steganographic Scheme}
\label{sec:amr}
In this section, we will propose a novel steganographic scheme,
which is called AMA,
to counter a targeted steganalyzer.
First, we will outline the basic idea of the proposed scheme.
Then we will discuss two most important operations in the proposed scheme, \textit{i.e.,} adversarial embedding and minimum alteration, in details.
Finally, we will give a practical implementation of AMA.

\subsection{Basic Idea}
\label{subsecIII:basic}
In the proposed scheme, the image elements are randomly divided into two groups, \textit{i.e.,}
a common group  containing common elements,
and an adjustable group containing \textit{adjustable elements}.
Data embedding is performed in two phases.
In the first phase, a portion of the stego message is embedded into the common group by using a conventional steganographic scheme.
In the second phase, the remaining part of the stego message is embedded into the adjustable group by using the proposed adversarial embedding scheme.
Adjustable elements are modified in such a way that
a targeted steganalyzer would output a wrong class label.
We use a well-known deep learning based steganalyzer, \textit{i.e.}, Xu's CNN \cite{XuJPEG},
as the targeted steganalyzer, since the gradient values of its loss function with respect to the input can be used
to guide the modification of adjustable elements.
Other steganalyzers possessing such a property may also be used.
The details will be given in Section \ref{subsecIII:advemb}.
In order to  prevent over-adapted to the targeted steganalyzer and enhance the security performance against other advanced steganalyzers, the number of adjustable elements is minimized, resulting in a minimization problem with constraints.
The details will be given in Section \ref{subsecIII:mr}.

\subsection{Adversarial Embedding}
\label{subsecIII:advemb}

Denote $y$  as the ground truth label of $\mathbf X$.
In steganalysis, we have $y \in \{0,1\}$, where  $0$ indicates a cover and  $1$ indicates a stego.
Let $L(\mathbf X,  y; \phi_{\mathcal{C,S}})$ be the loss function of a steganalyzer $\phi_{\mathcal{C,S}}$.
For example, for a deep neural network steganalyzer,
the binary decision could be given as
\begin{equation}\label{eq:binary_class}
  \phi_{\mathcal{C,S}}(\mathbf X)=
  \begin{aligned}
  \begin{cases}
    0,  \quad \text{if } F(\mathbf X) < 0.5, \\
    1,  \quad \text{if } F(\mathbf X) \ge 0.5,
    \end{cases}
  \end{aligned}
\end{equation}
where $F(\mathbf X) \in [0, 1]$ is the network output indicating the probability that $\mathbf X$ is a stego. The corresponding loss function may be designed in a form of cross entropy as
\begin{equation}\label{eq:Jfunction}
  L(\mathbf X,  y; \phi_{\mathcal{C,S}}) =
  -y \log \left( F(\mathbf X) \right) - (1-y) \log \left( 1 - F(\mathbf X) \right)
\end{equation}
In \cite{AD,deepFool,resistant}, adversarial examples are generated to fool ML models by updating input elements  $x_{i,j}$ according to the gradient of the loss function with respect to the input (abbreviated as \textit{gradient} if it is not specified otherwise), \textit{i.e.}, $\bigtriangledown_{x_{i,j}} L(\mathbf X, \hat{y}; \phi_{\mathcal{C,S}})$, by using a targeted label $\hat{y}$.
However, it is impossible to directly apply these methods for securing steganography.
In fact, modifying the elements of a stego image may lead to the failure of data extraction
thus contradicting the aim of steganography.
This motivates us to design an embedding method with two objectives of equal importance:
performing adversarial operation to combat steganalyzer $\phi_{\mathcal{C,S}}$
and performing data embedding to carry information.
To this end, we propose a method that we will call \textit{adversarial embedding} to generate
adversarial stego images
under the framework of steganographic distortion minimization \cite{simulator}.

In the distortion minimization framework,
steganography is formulated as an optimization problem with a payload constraint, \textit{i.e.},
\begin{equation}\label{equ:problem}
\min \limits_{\mathbf{S} } D(\mathbf{C}, \mathbf{S} ),  \quad
\text{s.t. }  \psi(\mathbf{S} ) = k,
\end{equation}
where $D(\mathbf{C}, \mathbf{S} )$ is a function measuring the
distortion caused by modifying $\mathbf{C}$ to $\mathbf{S}$, and
 $\psi(\mathbf S)$ represents the message payload extracted from $\mathbf S$ (measured in bits).
A typical additive distortion function for ternary embedding, such as those used in \cite{WOW,UNIWARD,HILL,UED,MP},
is defined as:
\begin{equation}\label{equ:add_distortion}
D({\bf C, \bf S}) = \sum_{i=1}^{H} \sum_{j=1}^{W} \rho_{i,j}^{+}\delta(  m_{i,j}-1 ) + \rho_{i,j}^{-}\delta(  m_{i,j}+1 ),
\end{equation}
where $m_{i,j}=s_{i,j}-c_{i,j}$ is the difference between the cover and the stego elements,
$\delta(\cdot)$ is an indication function:
\begin{equation}\label{equ:delta}
    \delta(x)=
    \begin{cases}
        1, \quad  x=0,\\
        0,  \quad otherwise,
    \end{cases}
\end{equation}
and
$\rho_{i,j}^{+}$ and $\rho_{i,j}^{-}$ are respectively the cost of increasing and decreasing $c_{i,j}$ by 1.
Although different steganographic schemes may employ different cost functions,
a rule of thumb is that large cost values are assigned to elements more likely to introduce abnormal artefacts leading to low probabilities of modification, and vice versa.
In most schemes, $\rho_{i,j}^{+}$ $=$ $\rho_{i,j}^{-}$,
leading to equal probabilities of increasing or decreasing $c_{i,j}$.
With the CMD (clustering modification direction) strategy \cite{CMDHILL},\cite{Denemark2015}, the costs of increasing or decreasing are asymmetrically updated during embedding
in favor to a synchronized direction in neighborhood.

In \cite{AD}, it is observed that
when a perturbation signal associated with a targeted label is added to the input,
the updated input, called \textit{adversarial example}, is usually misclassified into the targeted class by the ML classifier.
The perturbation signal can be designed in various ways, including using the gradient of the loss function with respect to the input.
Since adding a perturbation with the inverse sign of the gradient has an adversarial effect,
the objective of the proposed adversarial embedding is to modify image elements
in such a way that the sign of the modification tends to be in accordance with the inverse sign of the gradient.
To achieve such an objective with a high probability,
together with data embedding,
we operate under the distortion minimization framework by
defining the embedding costs as follows:
\begin{equation}
\rho_{i,j}^{+}
\begin{aligned}
\begin{cases}
\text{$<\rho_{i,j}^{-}$, \;if $\bigtriangledown_{x_{i,j}} L(\mathbf X, \hat{y}; \phi_{\mathcal{C,S}}) <0$},\\
\text{$=\rho_{i,j}^{-}$, \;if $\bigtriangledown_{x_{i,j}} L(\mathbf X, \hat{y}; \phi_{\mathcal{C,S}}) = 0$},\\
\text{$>\rho_{i,j}^{-}$, \;if $\bigtriangledown_{x_{i,j}} L(\mathbf X, \hat{y}; \phi_{\mathcal{C,S}}) >0$}.
\label{equ:rho adjust}
\end{cases}
\end{aligned}
\end{equation}
Such costs yield asymmetric probabilities of increasing and decreasing the element $x_{i,j}$, if the gradient is not zero.
In this way, data can be embedded into the image elements, and the direction of the modification has the effect of inducing the steganalyzer $\phi_{\mathcal{C,S}}$ to decide for the targeted label $\hat{y}=0$.

\subsection{Minimum Alteration of Adjustable Elements}
\label{subsecIII:mr}
With adversarial embedding, the adversarial stego images may effectively evade steganalysis.
However, since the costs of increasing and decreasing are asymmetric,
it increases the number of changed image elements.
The reason is that the maximum entropy can only be obtained when the image element has an equal probability of increasing and decreasing.
With the payload constraint, asymmetric costs lead to a higher change rate when compared to symmetric costs.
Although a higher change rate may not necessarily lead to a worse security performance,
we would still like to minimize it  by reducing the frequency of adversarial embedding.
This is due to three facts.
First, it is sufficient to fool the ML classifier
by using only a part of the elements to perform the adversarial operation, as shown in \cite{small}.
In fact, it is even unnecessary to perform adversarial embedding to those stego images
which are generated by conventional steganographic schemes but are already incorrectly classified by the imperfect steganalyzer.
Second, if all elements are used for adversarial embedding,
the generated adversarial stego images may be overly adapted to the targeted steganalyzer
and may possibly become more detectable by other advanced steganalyzers.
We may perform a minimum amount of alteration to prevent
introducing other detectable artefacts that can be exploited by an adversary-aware steganalyzer.
Third, when the change rate is minimized, the image quality should be preserved better.

We propose to divide image elements into two groups
\textit{i.e.}, a common group containing common elements for conventional steganographic embedding,
and an adjustable group containing adjustable elements for adversarial embedding.
The objective is that the amount of adjustable elements should be minimized
while the targeted steganalyzer should output a wrong class label.
Mathematically speaking, the problem is formulated as
\begin{equation}\label{equ:new_obj}
\begin{aligned}
\min \beta, \quad
\text{s.t. }  \phi_{\mathcal{C,S}}(\mathbf Z)=0 \text{ and } \psi(\mathbf Z)  = k,
\end{aligned}
\end{equation}
where $\beta \in [0, 1]$ denotes the ratio of the amounts of adjustable elements to all image elements.
It is obvious that there is no explicit solution to such a problem.
To solve it efficiently, the targeted steganalyzer is employed to numerically search for ``just enough'' amount of adjustable elements to satisfy the constraints in \eqref{equ:new_obj}.
The details will be described in the next subsection.

\subsection{A Practical Implementation of AMA}
\label{subsecIII:amr}
In this part, we present a practical AMA steganographic scheme.
Since JPEG images are widely used and pervasive on the Internet, we use them as cover.
We will use Xu-CNN \cite{XuJPEG} as the targeted steganalyzer and
J-UNIWARD \cite{UNIWARD} as the baseline steganographic scheme for conventional data embedding.
However, other image formats, steganalyzers, or conventional embedding schemes, may also be applicable,
as indicated in Section \ref{subsecIII:basic}.
The detailed steps of the proposed scheme are described as follows.
\begin{enumerate}

\item For a cover image $\mathbf C = (c_{i,j})^{H \times W }$, use a conventional cost function (such as in J-UNIWARD) to compute the initial embedding costs, \textit{i.e.},
    $\{\rho_{i,j}^{+}, \rho_{i,j}^{-} \}$, for the DCT coefficients. Initialize the parameter $\beta=0$.
\item Divide the elements in $\mathbf C$ into two disjoint groups,
    \textit{i.e.}, a common group containing
    $l_1 = [ H\times W\times (1-\beta) ]$ common elements, and an adjustable group containing
    $l_2 = H\times W - l_1$ adjustable elements.
    The positions of these two kinds of elements can be fixed in advance or randomized with the details of the randomization to be discussed later.
\item Embed $k_1 = [ k\times(1-\beta) ]$ bits into the common group using the initial embedding costs computed in Step 1 by applying a distortion minimization coding scheme, such as STC (syndrome-trellis codes) \cite{STCs}.
    The resulting image is denoted as $\mathbf Z_{c}$.

\item Compute the gradients $\bigtriangledown_{z_{i,j}} L(\mathbf Z_{c}, \hat{y}; \phi_{\mathcal{C,S}})$ of the steganalyzer using the targeted label $\hat{y}=0$.
    Update the embedding costs for the adjustable elements by
    \begin{equation}\label{eq:updatecost1}
     q_{i,j}^{+}=
        \begin{cases}
        {\rho_{i,j}^{+}}/{\alpha}, &
        \text{ if $\bigtriangledown_{z_{i,j}} L(\mathbf Z_{c}, 0; \phi_{\mathcal{C,S}})< 0$},    \\
        {\rho_{i,j}^{+}},  &
        \text{ if $\bigtriangledown_{z_{i,j}} L(\mathbf Z_{c}, 0; \phi_{\mathcal{C,S}})=0$},    \\
        {\rho_{i,j}^{+}}.{\alpha}, &
        \text{ if $\bigtriangledown_{z_{i,j}} L(\mathbf Z_{c}, 0; \phi_{\mathcal{C,S}})>0$},    \\
        \end{cases}
      \end{equation}
    \begin{equation}\label{eq:updatecost2}
     q_{i,j}^{-}=
        \begin{cases}
        {\rho_{i,j}^{-}}/{\alpha}, &
        \text{ if $\bigtriangledown_{z_{i,j}} L(\mathbf Z_{c}, 0; \phi_{\mathcal{C,S}})> 0$},    \\
        {\rho_{i,j}^{-}},  &
        \text{ if $\bigtriangledown_{z_{i,j}} L(\mathbf Z_{c}, 0; \phi_{\mathcal{C,S}})=0$},    \\
        {\rho_{i,j}^{-}}.{\alpha}, &
        \text{ if $\bigtriangledown_{z_{i,j}} L(\mathbf Z_{c}, 0; \phi_{\mathcal{C,S}})<0$},    \\
        \end{cases}
  \end{equation}
    where $\alpha$ is a scaling factor set to 2 in this work. Embed $k_2= k -k_1$ bits into the adjustable elements by using the updated embedding costs computed from \eqref{eq:updatecost1} and \eqref{eq:updatecost2} and the same coding scheme used for the common group.
    The resultant image is $\mathbf Z$.

\item Take $\mathbf Z$ as the input of the steganalyzer $\phi_{\mathcal{C,S}}$.
    If $\phi_{\mathcal{C,S}}(\mathbf Z)=0$, which means the adversarial stego $\mathbf Z$ can fool the steganalyzer with a minimum value of $\beta$,
    use $\mathbf Z$ as the output and terminate the embedding process.
    Otherwise, the amount of adjustable elements may not be enough.
    In this case, update $\beta$ by  $\beta + \Delta \beta$, and
    repeat Step 2 to Step 5 until $\beta=1$ .
    We use $\Delta \beta = 0.1$ in this work.
    If $\beta=1$ and $\phi_{\mathcal{C,S}}(\mathbf Z)=1$,
    which corresponds to the failure case of adversarial embedding,
    we just use a conventional steganographic scheme for embedding and output a conventional stego image.
\end{enumerate}

Since the same coding scheme, such as STC, is used both in the adjustable group and the common group,
the message receiver neither
needs to be informed about the value of $\beta$, nor needs to know which image elements belong to
to the adjustable group or the common group.
Data is extracted in the same way as the baseline steganographic scheme.

As we know, in most existing steganographic schemes,
an embedding order of image elements is generated by scrambling the indexes of image elements,
where the scrambling operation is determined by a secret key shared between the sender and the receiver.
The secret key can be fixed for different images, or changed as a session key.
In the AMA implementation,
the positions of the common elements and that of adjustable elements can be determined as follows.
First,
generate an embedding order in the same way
as the baseline steganographic scheme.
Then, the common group is formed
by the first $l_1 = [ H\times W\times (1-\beta) ]$ elements according to the embedding order.
Finally, the adjustable group is formed by the remaining elements.
In other words, the positions of adjustable elements
can be fixed or randomized for different images,
depending on whether the embedding order is fixed or randomized.

\section{Experiments}
\label{sec:exp}
In order to evaluate the performance of the proposed AMA scheme,
we conduct the following experiments.
\begin{enumerate}
  \item
  We evaluate the performance of AMA in the presence of an adversary-unaware steganalyst who
  trains his steganalyzer with conventional stego images.
  This corresponds to the most favorable case for the steganographer.
  It will be reported in Section \ref{subsecIV:unaware}
  \item
  We evaluate the performance of AMA in the presence of an adversary-aware steganalyst
  who re-trains his steganalyzer with adversarial stego images.
  This corresponds to the most challenging case for the steganographer.
  It will be reported in Section \ref{subsecIV:aware}
  \item
  We simulate the situation when the knowledge of the steganographer and that of the steganalyst are alternatively updated.
  To the best of our knowledge,
  this is the first work to investigate iterative adversarial conditions for steganography and steganalysis.
  It will be demonstrated in Section \ref{subsecIV:game}
  \item
  We show in Section \ref{subsecIV:factor} why adversarial embedding guided by gradients and minimum alteration are important in the proposed scheme.
  \item
  We discuss the role of randomizing the positions of the adjustable elements in Section \ref{subsecIV:discuss}.
  \item
  We perform some experiments on another image set for further evaluation in Section \ref{subsecIV:boss}.
\end{enumerate}

The common settings and notations in the experiments are described in Section \ref{subsecIV:setting}.
Some statistical information about the stego image sets is provided in Section \ref{subsecIV:stat}.

\begin{table*}[t!]
\renewcommand{\arraystretch}{1.5}
\caption{The security performance (in \%) against an adversary-unaware steganalyzer}
\label{tab:unaware}
\centering
\begin{tabular}{p{1.5cm}<{\centering} c  p{1.8cm}<{\centering}
p{0.2cm}p{0.25cm}p{0.2cm}p{0.01cm}
p{0.2cm}p{0.25cm}p{0.2cm}p{0.01cm}
p{0.2cm}p{0.25cm}p{0.2cm}p{0.01cm}
p{0.2cm}p{0.25cm}p{0.2cm}p{0.01cm}
p{0.2cm}p{0.25cm}p{0.2cm}p{0.01cm}
}
\toprule
& & &\multicolumn{3}{c}{\textbf {0.1 bpnzAC}}
& & \multicolumn{3}{c}{\textbf {0.2 bpnzAC}}
&&\multicolumn{3}{c}{\textbf {0.3 bpnzAC}}
&&\multicolumn{3}{c}{\textbf {0.4 bpnzAC}}
&&\multicolumn{3}{c}{\textbf {0.5 bpnzAC}}   \\
\midrule
\textbf{Steganalyzer} & \textbf{Steganography} & \textbf{Testing Set} &
{$\bm {P_{fa}}$} & $\bm {P_{md}}$ & $\bm {P_{e}}$ &&
{$\bm {P_{fa}}$} & $\bm {P_{md}}$ & $\bm {P_{e}}$ &&
{$\bm {P_{fa}}$} & $\bm {P_{md}}$ & $\bm {P_{e}}$ &&
{$\bm {P_{fa}}$} & $\bm {P_{md}}$ & $\bm {P_{e}}$ &&
{$\bm {P_{fa}}$} & $\bm {P_{md}}$ & $\bm {P_{e}}$&    \\
\midrule
\multirow{2}{*}{${ \phi_{\mathcal{C}_{B}^{0},\mathcal{S}_{B}^{0}} }$}
&{J-UNIWARD \cite{UNIWARD}}
& $\left\{ \mathcal{C}_{B}^{1tst}, \mathcal{S}_{B}^{1tst} \right\} $
&44.1 &42.3 &43.2 &
&32.5 &34.6 &33.6 &
&24.0 &24.8 &24.4 &
&17.5 &18.7 &18.1 &
&12.9 &13.4 &13.2  \\
&{Proposed AMA}
& $\left\{\mathcal{C}_{B}^{1tst}, \mathcal{Z}_{B}^{1tst} \right\}$
&44.1 &92.5 &68.3 &
&32.5 &98.6 &65.6 &
&24.0 &99.3 &61.6 &
&17.5 &99.6 &58.5 &
&12.9 &99.5 &56.2  \\
\midrule

\multirow{2}{*}{$   {\phi_{\mathcal{C}_{B}^{0},\mathcal{S}_{B}^{0}}^{\prime} }$}
&{J-UNIWARD \cite{UNIWARD}}
& $\left\{  \mathcal{C}_{B}^{1tst}, \mathcal{S}_{B}^{1tst} \right\}$
&47.7 &45.4 &46.5 &
&42.8 &40.4 &41.6 &
&36.7 &35.1 &35.9 &
&31.6 &29.1 &30.4 &
&25.7 &23.4 &24.6 & \\
&{Proposed AMA}
& $\left\{  \mathcal{C}_{B}^{1tst}, \mathcal{Z}_{B}^{1tst} \right\}$
&47.7 &47.1 &47.3 &
&42.8 &45.1 &43.9 &
&36.7 &43.2 &40.0 &
&31.6 &38.9 &35.3 &
&25.7 &36.2 &30.9 &\\
\midrule

\multirow{2}{*}{ $ \phi_{\mathcal{C}_{B}^{0},\mathcal{S}_{B}^{0}}^{\prime\prime}$ }
&{J-UNIWARD \cite{UNIWARD}}
& $\left\{  \mathcal{C}_{B}^{1tst}, \mathcal{S}_{B}^{1tst} \right\}$
&48.6 &47.4 &48.0 &
&45.3 &44.3 &44.8 &
&40.0 &41.6 &40.8 &
&36.0 &36.4 &36.2 &
&31.0 &30.7 &30.8\\
&{Proposed AMA}
& $\left\{\mathcal{C}_{B}^{1tst}, \mathcal{Z}_{B}^{1tst} \right\}$
&48.6 &47.9 &48.3 &
&45.3 &45.9 &45.5 &
&40.0 &44.8 &42.3 &
&36.0 &40.2 &38.1 &
&31.0 &35.3 &33.2 \\
\bottomrule

\end{tabular}
\end{table*}

\begin{table*}[t!]
\renewcommand{\arraystretch}{1.5}
\caption{The security performance (in \%) against an adversary-aware steganalyzer}
\label{tab:aware}
\centering
\begin{tabular}{p{1.5cm}<{\centering} c  p{1.8cm}<{\centering}
p{0.2cm}p{0.25cm}p{0.2cm}p{0.01cm}
p{0.2cm}p{0.25cm}p{0.2cm}p{0.01cm}
p{0.2cm}p{0.25cm}p{0.2cm}p{0.01cm}
p{0.2cm}p{0.25cm}p{0.2cm}p{0.01cm}
p{0.2cm}p{0.25cm}p{0.2cm}p{0.01cm}  }
\toprule
& & &\multicolumn{3}{c}{\textbf {0.1 bpnzAC}}
& & \multicolumn{3}{c}{\textbf {0.2 bpnzAC}}
&&\multicolumn{3}{c}{\textbf {0.3 bpnzAC}}
&&\multicolumn{3}{c}{\textbf {0.4 bpnzAC}}
&&\multicolumn{3}{c}{\textbf {0.5 bpnzAC}}   \\
\midrule
\textbf{Steganalyzer} &\textbf{Steganography} &\textbf{Testing Set} &
{$\bm {P_{fa}}$} & $\bm {P_{md}}$ & $\bm {P_{e}}$ &&
{$\bm {P_{fa}}$} & $\bm {P_{md}}$ & $\bm {P_{e}}$ &&
{$\bm {P_{fa}}$} & $\bm {P_{md}}$ & $\bm {P_{e}}$ &&
{$\bm {P_{fa}}$} & $\bm {P_{md}}$ & $\bm {P_{e}}$ &&
{$\bm {P_{fa}}$} & $\bm {P_{md}}$ & $\bm {P_{e}}$    \\
\midrule
{$ \phi_{\mathcal{C}_{B}^{1trn},\mathcal{S}_{B}^{1trn}}$}
&{J-UNIWARD \cite{UNIWARD}}
& $\left\{ \mathcal{C}_{B}^{1tst}, \mathcal{S}_{B}^{1tst} \right\}$
&45.8 &42.2 &44.0 &
&35.5 &32.5 &34.0 &
&25.0 &25.6 &25.3 &
&17.9 &19.6 &18.7 &
&13.1 &14.1 &13.6  \\
{$ \phi_{\mathcal{C}_{B}^{1trn},\mathcal{Z}_{B}^{1trn}} $}
&{Proposed AMA}& $\left\{\mathcal{C}_{B}^{1tst}, \mathcal{Z}_{B}^{1tst} \right\}$
&48.2 &46.8 &47.5 &
&39.4 &41.5 &40.4 &
&34.7 &32.5 &33.6 &
&27.4 &24.2 &25.8 &
&19.5 &18.9 &19.2  \\
\midrule

{$\phi_{\mathcal{C}_{B}^{1trn},\mathcal{S}_{B}^{1trn}}^{\prime}$}
&{J-UNIWARD \cite{UNIWARD}}& $\left\{\mathcal{C}_{B}^{1tst}, \mathcal{S}_{B}^{1tst} \right\}$
&48.4 &45.0 &46.7 &
&42.8 &40.5 &41.7 &
&37.1 &35.1 &36.1 &
&31.3 &29.5 &30.4 &
&25.3 &24.1 &24.7   \\
{$\phi_{\mathcal{C}_{B}^{1trn},\mathcal{Z}_{B}^{1trn}}^{\prime}$}
&{Proposed AMA}& $\left\{ \mathcal{C}_{B}^{1tst}, \mathcal{Z}_{B}^{1tst} \right\}$
&49.5 &45.6  &47.4 &
&47.2 &40.5  &43.7 &
&41.3 &37.5  &39.4 &
&36.1 &32.4  &34.2 &
&30.8 &27.8  &29.3 & \\
\midrule

{ $ \phi_{\mathcal{C}_{B}^{1trn},\mathcal{S}_{B}^{1trn}}^{\prime\prime}$ }
&{J-UNIWARD \cite{UNIWARD}}& $\left\{\mathcal{C}_{B}^{1tst}, \mathcal{S}_{B}^{1tst} \right\}$
&48.3 &47.7 &48.0 &
&45.1 &44.5 &44.8 &
&40.7 &41.0 &40.8 &
&36.6 &36.0 &36.3 &
&30.7 &31.0 &30.8   \\
{ $ \phi_{\mathcal{C}_{B}^{1trn},\mathcal{Z}_{B}^{1trn}}^{\prime\prime}$ }
&{Proposed AMA}& $\left\{\mathcal{C}_{B}^{1tst}, \mathcal{Z}_{B}^{1tst} \right\}$
&48.8 &47.8 &48.3 &
&46.9 &44.4 &45.7 &
&43.0 &41.2 &42.1 &
&38.7 &37.0 &37.9 &
&32.6 &32.5 &32.6  \\
\bottomrule

\end{tabular}
\end{table*}

\subsection{Settings}
\label{subsecIV:setting}

\subsubsection{Image set}
The following two cover image sets are respectively used.
\begin{itemize}
  \item Basic500k, denoted by $\mathcal{C}_{B}$.
  It is obtained by randomly selecting $5\times10^5$ JPEG images with size larger than 256$\times$256 from ImageNet and then cropping their left top 256$\times$256 regions. The images are further converted to grayscale and re-compressed into JPEG format with pquality factor 75. This dataset has been previously used in \cite{zeng} to train CNN steganalyzers.
  Unless specified otherwise, the experiments are carried out on this image set.
  To use the images efficiently  under different circumstances,
    $\mathcal{C}_{B}$ is randomly split into two disjoint subsets, $\mathcal{C}_{B}^{0}$ and $\mathcal{C}_{B}^{1}$, each with $2.5\times10^5$ images.

  \item JPEG-BOSSBase, denoted by $\mathcal{C}_{J}$.
  In order to verify the performance of AMA on an image set with distinct difference from $\mathcal{C}_{B}$,
  we generate this set with the resizing operation without any possible double JPEG compression artefacts.
  It is obtained based on
  $10000$ images from the public data set BOSSBase v1.01 \cite{BOSS}.
  The 512$\times$512 PGM format images are resized  to 256$\times$256 with a \textit{Lanczos2} resampling kernel, and then compressed into JPEG format with quality factor 75.
  The experiments in Section \ref{subsecIV:boss} are carried out on this image set.
  $\mathcal{C}_{J}$ is randomly split into two disjoint subsets, $\mathcal{C}_{J}^{0}$ and $\mathcal{C}_{J}^{1}$, each with $5000$ images.
\end{itemize}

\subsubsection{Steganalyzers}
Three different steganalyzers are used to evaluate the security of the steganographic schemes. The details are described as follows.
\begin{itemize}
  \item  Xu-CNN  steganalyzer \cite{XuJPEG}, denoted as $\phi^{}$.
  To the best of our knowledge, it is the best performing date-driven JPEG CNN steganalyzer.
  The 20-layer CNN steganalyzer was proposed by Xu, and we build the CNN structure and set all training parameters as in \cite{XuJPEG}, with the only difference that the batch size is set to 100 during the training stage, with 50 cover images and their corresponding stego counterparts.
  The CNN model trained at the $100000$\textit{-th} iteration is used as the steganalyzer.
  \item  GFR  steganalyzer \cite{GFR}, denoted as  $\phi^{\prime}$.
  It is based on 17000 histogram features generated with Gabor filters and an FLD ensemble classifier \cite{Ensemble}.
  \item  DCTR steganalyzer \cite{DCTR}, denoted as $\phi^{\prime\prime}$.
  It is based on 8000 dimensional DCT residual features and an FLD ensemble classifier \cite{Ensemble}.
\end{itemize}

The steganalytic performance is evaluated by the
missed detection rate as in \eqref{eq:md},
the false alarm rate as in \eqref{eq:fa}, and the total error rate as in \eqref{eq:te} .

\subsubsection{Steganographic schemes}
We use two steganographic schemes to generate stego images.
\begin{itemize}
 \item J-UNIWARD \cite{UNIWARD}: It is used as a baseline steganographic scheme. The embedding costs of DCT coefficients are calculated in the wavelet domain using a Daubechies wavelet filter bank.
    The corresponding stego image sets are denoted as
    $\mathcal{S}_{B}^{0}$, $\mathcal{S}_{B}^{1}$
    $\mathcal{S}_{J}^{0}$, and $\mathcal{S}_{J}^{1}$.
 \item AMA: In the proposed scheme,  J-UNIWARD is used to compute the initial embedding costs and perform the conventional embedding.
    The steganalyzer $\phi^{}_{\mathcal{C}_{B}^{0}, \mathcal{S}_{B}^{0}}$ based on Xu-CNN is used as the targeted steganalyzer for adversarial embedding.
    The corresponding adversarial stego image sets are denoted as
    $\mathcal{Z}_{B}^{0}$, $\mathcal{Z}_{B}^{1}$,
    $\mathcal{Z}_{J}^{0}$, and $\mathcal{Z}_{J}^{1}$.
    The scaling parameter used in \eqref{eq:updatecost1} and \eqref{eq:updatecost2} is set to $\alpha=2$, where we have tried $\alpha \in \{1.5, 2, 3, 5, 10 \}$ and found only minor difference in performance.
\end{itemize}

The optimal embedding simulator \cite{simulator} is employed for both J-UNIWARD and AMA.
The Matlab implementation
of J-UNIWARD is used.\footnote{It is downloaded from \url{http://dde.binghamton.edu/download/stego_algorithms/}}.
Our proposed AMA scheme is implemented using
TensorFlow with Python interface.
The experiments are run on a NVIDIA  Tesla K80 GPU platform.
The embedding payload is measured by bits per non-zero cover AC DCT coefficient (bpnzAC) as in \cite{UNIWARD,XuJPEG,zeng}.
In Section \ref{subsecIV:unaware} and \ref{subsecIV:aware},
we conduct experiments on 0.1, 0.2, 0.3, 0.4, and 0.5 bpnzAC.
For the rest of the experiments, we use 0.4 bpnzAC
since the steganalyzers perform better on higher payloads.

\subsection{Performance against an Adversary-unaware Steganalyst}
\label{subsecIV:unaware}
In this part, we  study the case where the knowledge of the steganalyzer is exposed to the steganographer, but the steganalyst is unaware of the adversarial operation and still use the current steganalyzer.
In particular, we assume
that the Xu-CNN steganalyzer $\phi_{\mathcal{C}_{B}^{0},\mathcal{S}_{B}^{0}}$, which has been trained on the image set $\left\{ \mathcal{C}_{B}^{0}, \mathcal{S}_{B}^{0} \right\}$,
is available to the steganographer.
Note that the steganographer does not necessarily need to have access to $\left\{ \mathcal{C}_{B}^{0}, \mathcal{S}_{B}^{0} \right\}$ given that the steganalyzer $\phi_{\mathcal{C}_{B}^{0},\mathcal{S}_{B}^{0}}$ is known.
The steganographer can
use $\phi_{\mathcal{C}_{B}^{0},\mathcal{S}_{B}^{0}}$
to generate an adversarial stego set $\mathcal{Z}_{B}^{1}$ from the cover set $\mathcal{C}_{B}^{1}$.
We would like to know how does the steganalyzer
$\phi_{\mathcal{C}_{B}^{0},\mathcal{S}_{B}^{0}}$
perform on classifying $\left\{ \mathcal{C}_{B}^{1tst}, \mathcal{Z}_{B}^{1tst} \right\}$
when compared to classifying  $\left\{ \mathcal{C}_{B}^{1tst}, \mathcal{S}_{B}^{1tst} \right\}$.
The experimental results are reported in  Table \ref{tab:unaware}.
Note that under the same payload rate,
the false alarm rate $P_{fa}$ is the same for $\left\{ \mathcal{C}_{B}^{1tst}, \mathcal{Z}_{B}^{1tst} \right\}$ and
$\left\{ \mathcal{C}_{B}^{1tst}, \mathcal{S}_{B}^{1tst} \right\}$,
due to the fact that the steganalyzer was trained on
$\left\{ \mathcal{C}_{B}^{0}, \mathcal{S}_{B}^{0} \right\}$ but tested on
$\mathcal{C}_{B}^{1tst}$, which is shared in
$\left\{ \mathcal{C}_{B}^{1tst}, \mathcal{Z}_{B}^{1tst} \right\}$ and
$\left\{ \mathcal{C}_{B}^{1tst}, \mathcal{S}_{B}^{1tst} \right\}$.
However, we can observe that the missed detection rate $P_{md}$  is much higher for
$\mathcal{Z}_{B}^{1tst}$ than for
$\mathcal{S}_{B}^{1tst}$.
These results indicate that the adversarial stego images generated by AMA can effective evade detection by the targeted steganalyzer.

In order to investigate
the case where the adversarial stego images are analyzed by steganalyzers other than the targeted one,
we conducted experiments by using two advanced steganalyzers, \textit{i.e., }
$\phi_{\mathcal{C}_{B}^{0},\mathcal{S}_{B}^{0}}^{\prime}$, and
$\phi_{\mathcal{C}_{B}^{0},\mathcal{S}_{B}^{0}}^{\prime\prime}$, to perform the same classification tasks.
The experimental results reported in Table \ref{tab:unaware}
show that the performance of these detectors on the adversarial stego images are, at least to some extent, worse than those obtained on the stego images generated by J-UNIWARD.
Although being designed to fool a targeted steganalyzer,
the AMA scheme shows a certain effectiveness also against non-targeted steganalyzers.
We speculate that this adaptability to other steganalyzers is due to the following facts.
1) The selected data-driven steganalyzer $\phi_{\mathcal{C}_{B}^{0},\mathcal{S}_{B}^{0}}$,
trained on an image set containing hundreds of thousands of  representative images,
is very powerful and have better detecting ability compared to other steganalyzers, hence 2) resisting such a powerful steganalyzer may implicitly preserve the statistics of the cover images, as shown in Section \ref{subsecIV:stat}, and therefore can also weaken other advanced steganalyzers (at least to some extent).

\subsection{Performance against an Adversary-aware Steganalyst}
\label{subsecIV:aware}
In this part, we study the case where the steganalyst is aware of the adversarial embedding operation.
As  stated in Section \ref{subsecII:steganalysis}, his best reaction is to re-train the steganalyzers with adversarial stego images.
We conducted experiments by dividing the cover image set
$\mathcal{C}_{B}^{1}$ into  $\mathcal{C}_{B}^{1trn}$
and $\mathcal{C}_{B}^{1tst}$, with $1.5\times10^5$ and $1\times10^5$ images, respectively.
The adversarial stego images $\mathcal{Z}_{B}^{1trn}$ and $\mathcal{Z}_{B}^{1tst}$ are generated as in Section \ref{subsecIV:unaware},
where the steganographer only relies on the steganalyzer $\phi_{\mathcal{C}_{B}^{0},\mathcal{S}_{B}^{0}}$
to generate adversarial stego images.
Then, we trained the steganalyzers based on $\left\{ \mathcal{C}_{B}^{1trn}, \mathcal{Z}_{B}^{1trn} \right\}$
and tested on $\left\{ \mathcal{C}_{B}^{1tst}, \mathcal{Z}_{B}^{1tst} \right\}$.
In this way, the image sets for data embedding (\textit{i.e.,} $\mathcal{C}_{B}^{1trn}$ and $\mathcal{C}_{B}^{1tst}$) and that for the training targeted steganalyzer (\textit{i.e.,} $\mathcal{C}_{B}^{0}$) are different, thus ensuring that
AMA does not use any prior knowledge of the image set.

The experimental results we obtained are reported in Table \ref{tab:aware}.
It can be observed that compared to the targeted steganalyzer, which is easily fooled by the adversarial stego images,
a re-trained steganalyzer can better detect the adversarial embedding operations.
However, compared to the baseline J-UNIWARD scheme,
the proposed AMA scheme still achieves a better security performance.
For example, AMA gets a $25.8\%$ total error rate for 0.4 bpnzAC, which is comparable to
J-UNIWARD with $25.3\%$ for 0.3 bpnzAC.
This means that under the same risk level of detection, AMA attains 0.1 bpnzAC more payload.
As also shown in Table \ref{tab:aware}, when we use the other two non-targeted steganalyzers
$\phi^{\prime}$ and $\phi^{\prime\prime}$ for detection,
higher total error rates are obtained on
$\left\{ \mathcal{C}_{B}^{1tst}, \mathcal{Z}_{B}^{1tst} \right\}$
than on $\left\{ \mathcal{C}_{B}^{1tst}, \mathcal{S}_{B}^{1tst} \right\}$, showing, once again, that AMA outperforms the baseline scheme.

\begin{table}[t!]
\caption{The security performance (in \%) of the iterative game when the steganalyzer evolved alternatively between adversary-unaware and adversary-aware.}
\label{tab:game}
\centering
\begin{tabular}{p{0.5cm}<{\centering} c p{1.5cm}<{\centering} ccc}
\toprule
\textbf{Round} &
\textbf{Testing set}&
\textbf{Steganalyzer} &
$\bm {P_{fa}}$ &
$\bm {P_{md}}$ &
$\bm {P_{e}}$
\\
\midrule
\multirow{2}{*}{1} &
\multirow{2}{*}{ $\{\mathcal{C}_{B}^{1tst},\mathcal{Z}_{B}^{1tst} \}$ }&
$\phi_{\mathcal{C}_{B}^{0},\mathcal{S}_{B}^{0}}$ &17.5 &99.6 &58.5 \\
& & $\phi_{\mathcal{C}_{B}^{1trn},\mathcal{Z}_{B}^{1trn}}$ &27.4 &24.2 &25.8 \\
\midrule
\multirow{2}{*}{2} &
\multirow{2}{*}{ $\{\mathcal{C}_{B}^{1tst},\mathcal{\dot{Z}}_{B}^{1tst} \}$ }&
$\phi_{\mathcal{C}_{B}^{0},\mathcal{Z}_{B}^{0}}$
&24.0 &98.3 &61.2 \\
& & $\phi_{\mathcal{C}_{B}^{1trn},\mathcal{\dot{Z}}_{B}^{1trn}}$ &26.3 &26.2& 26.3 \\
\midrule
\multirow{2}{*}{3} &
\multirow{2}{*}{ $\{\mathcal{C}_{B}^{1tst},\mathcal{\ddot{Z}}_{B}^{1tst} \}$ }&
$\phi_{\mathcal{C}_{B}^{0},\mathcal{\dot{Z}}_{B}^{0}}$
&26.9 &97.6 &62.3 \\
& & $\phi_{\mathcal{C}_{B}^{1trn},\mathcal{\ddot{Z}}_{B}^{1trn}}$ &24.5 &26.1 &25.3\\
\midrule
\multirow{2}{*}{4} &
\multirow{2}{*}{ $\{\mathcal{C}_{B}^{1tst},\mathcal{\dddot{Z}}_{B}^{1tst} \}$ }&
$\phi_{\mathcal{C}_{B}^{0},\mathcal{\ddot{Z}}_{B}^{0}}$
&24.2 &98.3 &61.3 \\
& & $\phi_{\mathcal{C}_{B}^{1trn},\mathcal{\dddot{Z}}_{B}^{1trn}}$ &23.9 &23.8& 23.8 \\
\midrule
\multirow{2}{*}{5} &
\multirow{2}{*}{ $\{\mathcal{C}_{B}^{1tst},\mathcal{\ddddot{Z}}_{B}^{1tst} \}$ }&
$\phi_{\mathcal{C}_{B}^{0},\mathcal{\dddot{Z}}_{B}^{0}}$ &23.9 &99.0 &61.5 \\
& & $\phi_{\mathcal{C}_{B}^{1trn},\mathcal{\ddddot{Z}}_{B}^{1trn}}$ &23.5 &22.9& 23.2 \\
\bottomrule
\end{tabular}
\end{table}

\subsection{Sequential Iterative Game between Steganographer and Steganalyst}
\label{subsecIV:game}
In this part, we study a scenario wherein the steganographer and the steganalyst
adjust their strategies iteratively each time by adapting their knowledge about the scheme adopted by the adversary.
This process can be simulated by performing experiments similarly to Section \ref{subsecIV:unaware} and Section \ref{subsecIV:aware} iteratively using the Xu-CNN steganalyzer.
To define the iterations, we make the following assumptions for each round.
\begin{itemize}
  \item
  The adversary-unaware steganalyst
  is unaware of the adversarial stego images generated \textit{in the current round}.
  For the first round, plain stego images generated with the baseline steganographic scheme are used for training.
  For the subsequent rounds, adversarial stego images, which are generated in the same way as the steganographer \textit{in the previous round}, are used.
  The steganalyzer is trained on $\mathcal{C}_{B}^{0}$ and its stego (or adversary stego) counterpart.
  \item The steganographer sets the targeted steganalyzer to be the same as the adversary-unaware steganalyst \textit{in the current round} and tries to attack it by generating adversarial stego images from $\mathcal{C}_{B}^{1}$.
  \item  The adversary-aware steganalyst is aware of the adversarial operation performed \textit{in the current round}.
   He trains the classifier based on $\mathcal{C}_{B}^{1trn}$ and the adversarial stego counterpart in the current round.
  \item To ease the comparison, the $\mathcal{C}_{B}^{1tst}$ and its corresponding adversarial stego counterpart are used to evaluate the performance for both the adversary-unaware steganalyzer and the adversary-aware steganalyzer.
\end{itemize}
Although the iterative process can be endless, we performed five rounds of iterations to illustrate the dynamic effects.
The adversary-unaware steganalyst uses J-UNIWARD to generate the plain stego image set $\mathcal{S}$ in the first round. The steganographer generates the adversarial stego sets $\mathcal{Z}$, $\mathcal{\dot{Z}}$, $\mathcal{\ddot{Z}}$, $\mathcal{\dddot{Z}}$, and $\mathcal{\ddddot{Z}}$ from the first to the fifth round, respectively.
The embedding payload is set to 0.4 bpnzAC.
According to the results shown in  Table \ref{tab:game},
we can make the following observations.
\begin{enumerate}
  \item In the same round, the adversary-unaware steganalyzer has
        a higher total error rate than the adversary-aware steganalyzer, mainly due to its higher missed detection rate.
        This implies that
        the steganographer can always effectively fool the targeted steganalyzer,
        while the adversary-aware steganalyst effectively exploits the knowledge about the adversary's operations.
  \item As the iteration goes on, although the gap in missed detection rate between the adversary-unaware steganalyzer and the adversary-aware steganalyzer fluctuates (\textit{i.e.,} 75.4\%, 72.1\%, 71.5\%, 74.5\%, 76.1\% for the five rounds),
        the gap in false alarm rate narrows (\textit{i.e.,} 9.9\%, 2.3\%, 2.4\%, 0.3\%, 0.4\% for the five rounds).
        The gap in total error rate between the two steganalyzers widens (\textit{i.e.,} 32.7\%, 34.9\%, 37\%, 37.5\%, 38.3\% for the five rounds).
\end{enumerate}

We observe that when an adversary-aware steganalyzer is used, the steganographer is, in fact, in an adversary-unaware situation. Conversely, in the case of an adversary-unaware steganalyzer, the steganographer is aware of the actions of his adversary.
Since the steganographer and the steganalyzer
alternatively take move in the iterative experiments,
the results indicate that
the player who has the adversary information and makes the last move has a great advantage.

\subsection{Investigation on Two Important Components in AMA}
\label{subsecIV:factor}

Performing adversarial embedding according to the inverse signs of gradients and using minimum alteration
are the two most important components
of the AMA scheme,
we then conducted some experiments to investigate the effectiveness of each component.
Both adversary-unaware and adversary-aware CNN steganalyzers are used for the evaluation, and the embedding payload is set to 0.4 bpnzAC.

\subsubsection{Case I: reversing adversarial embedding operation}
In the AMA scheme, the embedding costs of adjustable elements are asymmetrically adjusted according to the inverse signs of the gradients,
as shown in \eqref{eq:updatecost1} and \eqref{eq:updatecost2}.
Now, we use the signs of the gradients, instead of the inverse signs, as in the following equations, to perform a comparative experiment:
    \begin{equation}\label{eq:updatecost3}
     q_{i,j}^{+}=
        \begin{cases}
        {\rho_{i,j}^{+}}/{\alpha}, &
        \text{ if $\bigtriangledown_{z_{i,j}} L(\mathbf Z_{c}, 0; \phi_{\mathcal{C,S}})> 0$},    \\
        {\rho_{i,j}^{+}},  &
        \text{ if $\bigtriangledown_{z_{i,j}} L(\mathbf Z_{c}, 0; \phi_{\mathcal{C,S}})=0$},    \\
        {\rho_{i,j}^{+}}.{\alpha}, &
        \text{ if $\bigtriangledown_{z_{i,j}} L(\mathbf Z_{c}, 0; \phi_{\mathcal{C,S}})<0$},    \\
        \end{cases}
      \end{equation}
    \begin{equation}\label{eq:updatecost4}
     q_{i,j}^{-}=
        \begin{cases}
        {\rho_{i,j}^{-}}/{\alpha}, &
        \text{ if $\bigtriangledown_{z_{i,j}} L(\mathbf Z_{c}, 0; \phi_{\mathcal{C,S}})< 0$},    \\
        {\rho_{i,j}^{-}},  &
        \text{ if $\bigtriangledown_{z_{i,j}} L(\mathbf Z_{c}, 0; \phi_{\mathcal{C,S}})=0$},    \\
        {\rho_{i,j}^{-}}.{\alpha}, &
        \text{ if $\bigtriangledown_{z_{i,j}} L(\mathbf Z_{c}, 0; \phi_{\mathcal{C,S}})>0$}.    \\
        \end{cases}
  \end{equation}
The results are shown in Table \ref{tab:factor}.
Compared with the previous results (see Table \ref{tab:unaware} and \ref{tab:aware}),
the total error rate of the adversary-unaware steganalyzer drops from 58.5\% to 21.6\%,
and that of the adversary-aware steganalyzer  from 25.8\% to 19.3\%.
The degraded performance
indicates that taking into account the signs of the gradients plays an important role in producing the adversarial effect.

\subsubsection{Case II: disabling minimum alteration}
In the AMA scheme, the number of adjustable elements is minimized through iteratively finding a minimum value of $\beta$
for \eqref{equ:new_obj}.
In the comparative experiment, we use a fixed value of $\beta$ for each image,
and thus the amount of adjustable elements is the same
for all the images.
The results we have got for $\beta=0.1$, $0.3$, and $0.5$ are presented in Table \ref{tab:factor}.
It can be observed that as $\beta$ increases,
the missed detection rate of the adversary-unaware steganalyzer increases,
but the total error rate of the adversary-aware steganalyzer decreases.
The results indicate that when increasing the number of adjustable elements,
it becomes easier to fool the targeted steganalyzer. However,
an excess of adversarial operations may introduce unnecessary artefacts, leading to easier detection by
an adversary-aware steganalyzer.
Consequently, it is a better choice to use ``just enough'' amount of adjustable elements by balancing the performance of
an adversary-unaware steganalyzer and an adversary-aware steganalyzer.

\begin{table*}[t!]
\caption{The security performance (in \%) with different setting for AMA under the payload of 0.4 bpnzac}
\label{tab:factor}
\centering
\begin{tabular}{c c
p{0.3cm}p{0.3cm}p{0.3cm}p{0.05cm}
p{0.3cm}p{0.3cm}p{0.3cm}p{0.05cm}
p{0.3cm}p{0.3cm}p{0.3cm}p{0.05cm}
p{0.3cm}p{0.3cm}p{0.3cm}p{0.05cm} }
\toprule
&&\multicolumn{3}{c}{\textbf {Case I}}
&&\multicolumn{3}{c}{\textbf {Case II ($\bm{\beta=0.1$})}}
&&\multicolumn{3}{c}{\textbf {Case II ($\bm{\beta=0.3$})}}
&&\multicolumn{3}{c}{\textbf {Case II ($\bm{\beta=0.5$})}}\\
\midrule
\textbf{Steganalyzer} &\textbf{Testing Set} &
{$\bm {P_{fa}}$}  & $\bm {P_{md}}$  & $\bm {P_{e}}$ &&
{$\bm {P_{fa}}$}  & $\bm {P_{md}}$  & $\bm {P_{e}}$ &&
{$\bm {P_{fa}}$}  & $\bm {P_{md}}$  & $\bm {P_{e}}$ &&
{$\bm {P_{fa}}$}  & $\bm {P_{md}}$  & $\bm {P_{e}}$ \\
\midrule
$\phi_{\mathcal{C}_{B}^{0},\mathcal{S}_{B}^{0}}$
& $\left\{ \mathcal{C}_{B}^{1tst}, \mathcal{Z}_{B}^{1tst} \right\}$
&17.5 &25.7 &21.6 &
&17.5 &49.2 &33.3 &
&17.5 &87.0 &52.3 &
&17.5 &96.2 &56.9 \\
\midrule
$\phi_{\mathcal{C}_{B}^{1trn},\mathcal{Z}_{B}^{1trn}}$
& $\left\{ \mathcal{C}_{B}^{1tst}, \mathcal{Z}_{B}^{1tst} \right\}$
&18.2 &20.5 &19.3 &
&23.9 &22.8 &23.3 &
&23.0 &22.7 &22.8 &
&18.1 &19.4 &18.7 \\
\bottomrule
\end{tabular}
\end{table*}

\subsection{Supplementary Statistical Information}
\label{subsecIV:stat}
To further investigate the proposed AMA scheme,
we provide some supplementary statistical information on the adversarial stego images as follows.
\subsubsection{Frequency of adversarial embedding operation}
To investigate the statistics on how many adjustable elements are used in the AMA scheme,
the occurrences of $\beta$ in generating the $2.5\times10^5$ adversarial stego images
$\mathcal{Z}_{B}^{1}$ are given in Table \ref{tab:beta}.
Based on the statistics, we can make the following observations.
\begin{itemize}
  \item For a low payload, such as 0.1 bpnzAC, since the steganalyzer is less effective in detecting plain stego images,
        adversarial embedding is not necessary for a large portion of the stego images, which corresponds to the case of $\beta=0$.
        As the payload increases, more stego images requires adversarial embedding ($\beta\neq0$).
  \item A lower failure rate of adversary embedding is obtained for a higher payload
        (from 7.52\% on 0.1 bpnzAC to 0.47\% on 0.5 bpnzAC). This is due to the fact that
        more elements are involved in modification
        as the payload increase.
        For instance, less than 2\% elements are used for modification for 0.1 bpnzAC, while more than 11\% elements are used for modification for 0.5 bpnzAC, as shown in Table \ref{tab:changerate}.
        Note that the failure rate is exactly the same as $(1-P_{md})$ of the adversary-unaware CNN steganalyzer given in Table \ref{tab:unaware}.
  \item For all payloads, larger values of $\beta$ occur less frequently than lower values.
        However, this phenomenon cannot be taken for granted since
        it may be due to the specific images, the baseline steganographic scheme,
        the targeted steganalyzer,
        and the step $\Delta \beta$ used to search the minimum $\beta$.
\end{itemize}

\subsubsection{Modification rate}
In Section \ref{subsecIII:mr},
we have stated that
adversarial embedding would lead to an increasing number of modified image elements
due to the asymmetric costs assigned to the adjustable elements.
We define the modification rate as the ratio of the number of changed coefficients to the total amount of non-zero AC DCT coefficients.
In Table \ref{tab:changerate}, we show the
averaged modification rate
for J-UNIWARD and AMA under different payloads on the image set $\mathcal{C}_{B}^{1}$.
As expected, we can observe that
the modification rates for AMA are slightly higher than for J-UNIWARD.
Besides, the gap in the modification rate between J-UNIWARD and AMA widens
as the payload increases (0.04\%, 0.07\%, 0.11\%, 0.15\%, 0.2\% for the five payloads, respectively).
This is due to the fact that more cases of $\beta\neq 0$ occur for a higher payload, as indicated in Table \ref{tab:beta}.

\subsubsection{Feature distance}
\label{subsecIV:distance}
In Section \ref{subsecIV:unaware},
we claim that the good performance of AMA
against statistical feature-based steganalyzers may be due to an implicit ability in
preserving some statistics.
To verify this statement, we use the Maximum Mean Discrepancy (MMD)\footnote{The MMD toolbox can be downloaded from \url{http://dde.binghamton.edu/tomas/mmdToolBox.zip}.
} \cite{Pevny2008}
between the cover image set and the stego image set
on the feature space formed by 17000-D GFR features \cite{GFR}.
Since it requires extremely large memories
for computing all $2.5\times10^5$ images in $\mathcal{C}_{B}^{1}$, we randomly select
10000 images.
The same scaling factor is used for both J-UNIWARD and AMA.
The results are shown in Table \ref{tab:mmd}.
It can be observed that AMA has a lower MMD value than J-UNIWARD under the same payload, indicating it preserves the GFR features better,
even though its modification rate is higher.

\begin{table}[t!]
\caption{The frequencies of occurrences of $\beta$ (in \%) in generating stego image set $\mathcal{Z}_{B}^{1}$ for each payload.
The sum of each column is 100\%. }
\label{tab:beta}
\centering
\begin{tabular}{ cccccc}
\toprule
\multirow{2}{*}{$\bm \beta$}&
\textbf{0.1}  &
\textbf{0.2}  &
\textbf{0.3}  &
\textbf{0.4}  &
\textbf{0.5}
\\
&
\textbf{bpnzAC}  &
\textbf{bpnzAC}  &
\textbf{bpnzAC}  &
\textbf{bpnzAC}  &
\textbf{bpnzAC}
\\
\midrule
\textbf{0}   &40.61 &34.08 &24.74 &18.60 &13.35 \\
\textbf{0.1} &13.31 &22.00 &28.15 &31.67 &32.71 \\
\textbf{0.2} &9.83  &16.65 &22.56 &26.21 &28.08 \\
\textbf{0.3} &7.85  &11.00 &12.70 &13.47 &14.73 \\
\textbf{0.4} &6.13  &6.50  &5.98  &5.68  &6.39 \\
\textbf{0.5} &4.70  &3.79  &2.71  &2.33  &2.56 \\
\textbf{0.6} &3.62  &2.16  &1.27  &0.95  &1.02 \\
\textbf{0.7} &2.78  &1.26  &0.66  &0.40  &0.40 \\
\textbf{0.8} &2.09  &0.76  &0.35  &0.22  &0.19 \\
\textbf{0.9} &1.50  &0.43  &0.20  &0.10  &0.09 \\
\textbf{1} &0.06  &0.02  &0.01  &0.01  &0.01 \\
\textbf{fail}    &7.52  &1.35  &0.67  &0.36  &0.47 \\
\bottomrule
\end{tabular}
\end{table}

\begin{table}[t!]
\renewcommand{\arraystretch}{1}
\caption{The modification rate computed as the change per non-zero AC DCT coefficient (in \%) for the two steganographic schemes under different payloads.}
\label{tab:changerate}
\centering
\begin{tabular}{c p{0.8cm}p{0.8cm}p{0.8cm}p{0.8cm}p{0.8cm}}
\toprule
\multirow{2}{*}{\textbf{Steganography}}&
\textbf{0.1}  &
\textbf{0.2}  &
\textbf{0.3}  &
\textbf{0.4}  &
\textbf{0.5}
\\
&
\textbf{bpnzAC}  &
\textbf{bpnzAC}  &
\textbf{bpnzAC}  &
\textbf{bpnzAC}  &
\textbf{bpnzAC}
\\

\midrule
{J-UNIWARD \cite{UNIWARD}}&
1.80 &
3.97 &
6.32 &
8.80 &
11.37
\\
{Proposed AMA}&
1.84 &
4.04 &
6.43 &
8.95 &
11.57
\\

\bottomrule
\end{tabular}
\end{table}

\begin{table}[t!]
\renewcommand{\arraystretch}{1}
\caption{The MMD (in $1\times 10^{-7}$) for the two steganographic schemes under different payloads.}
\label{tab:mmd}
\centering
\begin{tabular}{c p{0.8cm}p{0.8cm}p{0.8cm}p{0.8cm}p{0.8cm}}
\toprule
\multirow{2}{*}{\textbf{Steganography}}&
\textbf{0.1}  &
\textbf{0.2}  &
\textbf{0.3}  &
\textbf{0.4}  &
\textbf{0.5}
\\
&
\textbf{bpnzAC}  &
\textbf{bpnzAC}  &
\textbf{bpnzAC}  &
\textbf{bpnzAC}  &
\textbf{bpnzAC}
\\

\midrule
{J-UNIWARD \cite{UNIWARD}}&
3.6&
27.6&
95.6&
239.7&
499.1
\\
{Proposed AMA}&
2.5 &
19.2 &
70.4 &
189.3 &
409.8
\\
\bottomrule
\end{tabular}
\end{table}

\subsection{Discussion on the Role of Randomizing the Positions of Adjustable Elements}
\label{subsecIV:discuss}

In our previous experiments, the positions of adjustable elements are randomized by using different embedding orders for different images.
One question is whether there is a difference in
security performance between randomized
positions and fixed positions.
In order to investigate the role of randomizing the positions of adjustable elements, in the following we report the results of two comparative experiments.

In the first experiment, we use a fixed embedding order for different images.
As indicated in Section \ref{subsecIII:amr}, the fixed embedding order results in the fixed positions of adjustable elements.
We adopt the same setting we have used in Section \ref{subsecIV:unaware} and \ref{subsecIV:aware}. Adversary-unaware and adversary-aware CNN based steganalyzers are respectively used for detection.
The results we have got are shown in Table \ref{tab:constant}.
The improved, or deteriorated, of the performance with respect to the implementation adopting the randomized positions are shown in the parenthesis in the table.
It can be observed that AMA with the fixed positions of adjustable elements and that with the randomized positions of adjustable elements do not have obvious difference in performance against the CNN based steganalyzers.

As a second experiment, we use a fixed embedding order and a fixed number of adjustable elements ($\beta=0.3$) for each image.
The payload is set to 0.4 bpnzAC.
The results we have got are given in Table \ref{tab:constantbeta}.
The comparison with the implementation using a randomized embedding order
(Case II  with $\beta=0.3$ in Table \ref{tab:factor}),
is shown in the parenthesis.
It can be observed that
the performance does not change much for an adversary-unaware steganalyzer,
while it degrades greatly for an adversary-aware steganalyzer.
This phenomenon is interesting.
Although the fixed positions of adjustable elements are not directly leaked to the adversary-aware steganalyzer,
the experimental evidence shows that the data-driven steganalyzer can automatically learn such information.
In a similar scenario, when the same key is re-used for data embedding simulation, a CNN based method \cite{Pibre2016} is highly effective in detecting
different stego images with synchronized modification locations.
The performance drops greatly when different keys are used for different images.
The phenomenon does not occur for feature based steganalyzers. We speculate that modifications in the same location may present a chance of ``collision attack''
from the perspective of CNN based steganalyzers.
The neurons may learn strong activations from the synchronized embedding locations.
Since AMA employs minimum alteration, the collision effect is eliminated,
even when a fixed embedding order is used, as the results reported in Table \ref{tab:constant} show.

Based on the previous discussion, in order to improve the security of the stego images, a steganographer may want to use a random embedding order.
However, this may require to transmit a secret key from the sender to the receiver.
Some previous works propose to establish a secret channel for sharing this and other kinds of side information, or to embed the side information in the stego media \cite{luo2010edge}.
Another possibility is to compute a robust image hash and use the hash value as a secret key which can be
extracted by the receiver.
Since the random embedding order does not play an important role in the security of AMA,
we do not discuss its implementation in this work.

\subsection{Performance on JPEG-BOSSBase Image Set}
\label{subsecIV:boss}
In this part, we utilize the image set
JPEG-BOSSBase to further evaluate the performance of AMA.
The Xu-CNN  steganalyzer $\phi^{}_{\mathcal{C}_{B}^{0}, \mathcal{S}_{B}^{0}}$ trained on
Basic500k is still used as the targeted steganalyzer in the AMA scheme.
We use three adversary-aware steganalyzers to detect AMA,
and use J-UNIWARD as the baseline for comparison.
The embedding payload is set to 0.4 bpnzAC.
From the results shown in Table \ref{tab:boss},
we can observe that AMA performs better than J-UNIWARD on JPEG-BOSSBase.
The results indicate that the good performance of the proposed AMA scheme does not rely much on a specific image set.

\begin{table}[t!]
\renewcommand{\arraystretch}{1}
\caption{The security performance (in \%) of AMA
with a fixed embedding order against the adversary-unaware steganalyzer and the adversary-aware steganalyzer.
The testing image set is $\left\{\mathcal{C}_{B}^{1tst}, \mathcal{Z}_{B}^{1tst} \right\}$.
Performance comparison with the implementation using a randomized embedding order is shown in the parenthesis.}
\label{tab:constant}
\centering
\begin{tabular}{p{1.2cm} ccccc}
\toprule
\multirow{2}{*}{\textbf{Steganalyzer}}&
\textbf{0.1}  &
\textbf{0.2}  &
\textbf{0.3}  &
\textbf{0.4}  &
\textbf{0.5}
\\
&
\textbf{bpnzAC}  &
\textbf{bpnzAC}  &
\textbf{bpnzAC}  &
\textbf{bpnzAC}  &
\textbf{bpnzAC}
\\
\midrule
\multirow{2}{*}{$ \phi_{\mathcal{C}_{B}^{0},\mathcal{S}_{B}^{0}}$} &
68.32 &
65.61 &
61.68 &
58.58 &
56.24  \\
&
($\downarrow$0.02)&
($\downarrow$0.02)&
($\uparrow$0.04)&
($\uparrow$0.04) &
($\uparrow$0.01)\\
\midrule
\multirow{2}{*}{$ \phi_{\mathcal{C}_{B}^{1trn},\mathcal{Z}_{B}^{1trn}} $} &
47.49 &
40.60 &
33.05 &
25.18 &
19.05 \\
&
($\downarrow$0.05) &
($\uparrow$0.15) &
($\downarrow$0.58) &
($\downarrow$0.65) &
($\downarrow$0.12) \\
\bottomrule
\end{tabular}
\end{table}

\begin{table}[t!]
\caption{The security performance (in \%) of AMA
with a fixed embedding order
and a fixed number of adjustable elements ($\beta=0.3$) against the adversary-unaware steganalyzer and the adversary-aware steganalyzer.
The testing image set is $\left\{\mathcal{C}_{B}^{1tst}, \mathcal{Z}_{B}^{1tst} \right\}$.
Performance comparison with the implementation using a randomized embedding order is shown in the parenthesis.}
\label{tab:constantbeta}
\centering
\begin{tabular}{c p{0.3cm}p{0.3cm}p{0.05cm}
p{0.3cm}p{0.3cm}p{0.05cm}
p{0.3cm}p{0.3cm}p{0.05cm}}
\toprule
\textbf{Steganalyzer}
&\multicolumn{3}{c}{$\bm{P_{fa}}$}
&\multicolumn{3}{c}{$\bm{P_{md}}$}
&\multicolumn{3}{c}{$\bm{P_{e}}$} \\
\midrule
$\phi_{\mathcal{C}_{B}^{0},\mathcal{S}_{B}^{0}}$
&17.5 &(-) &
&87.2 &($\uparrow$0.2)&
&52.4 &($\downarrow$0.1)&
\\
$ \phi_{\mathcal{C}_{B}^{1trn},\mathcal{Z}_{B}^{1trn}}$
&14.1 &($\downarrow$8.9)&
&15.1 &($\downarrow$7.6)&
&14.6 &($\downarrow$8.2)&
\\
\bottomrule
\end{tabular}
\end{table}

\begin{table}[t!]
\caption{The security performance (in \%) on JPEG-BOSSBase image set under the payload of 0.4 bpnzAC}
\label{tab:boss}
\centering
\begin{tabular}{p{1.2cm}<{\centering} c p{1.4cm}<{\centering} ccc}
\toprule
\textbf{Steganalyzer}
&\textbf{Steganography}
&\textbf{Testing Set}
&$\bm{P_{fa}}$
&$\bm{P_{md}}$
&$\bm{P_{e}}$ \\
\midrule
$\phi_{\mathcal{C}_{J}^{0},\mathcal{S}_{J}^{0}}$
&J-UNIWARD \cite{UNIWARD}
&$\left\{ \mathcal{C}_{J}^{1}, \mathcal{S}_{J}^{1} \right\}$
&15.4 &23.6 &19.5    \\
$\phi_{\mathcal{C}_{J}^{0},\mathcal{Z}_{J}^{0}}$
&Proposed  AMA
& $\left\{ \mathcal{C}_{J}^{1}, \mathcal{Z}_{J}^{1} \right\}$
&21.8 &31.1 &26.4  \\
\midrule
$\phi_{\mathcal{C}_{J}^{0},\mathcal{S}_{J}^{0}}^{\prime}$
&J-UNIWARD \cite{UNIWARD}
&$\left\{ \mathcal{C}_{J}^{1}, \mathcal{S}_{J}^{1} \right\}$
&21.8 &21.2 &21.5    \\
$\phi_{\mathcal{C}_{J}^{0},\mathcal{Z}_{J}^{0}}^{\prime}$
& Proposed AMA
& $\left\{ \mathcal{C}_{J}^{1}, \mathcal{Z}_{J}^{1} \right\}$
&25.1 &24.7 &24.9  \\
\midrule
$\phi_{\mathcal{C}_{J}^{0},\mathcal{S}_{J}^{0}}^{\prime\prime}$
&J-UNIWARD \cite{UNIWARD}
&$\left\{ \mathcal{C}_{J}^{1}, \mathcal{S}_{J}^{1} \right\}$
&28.5 &28.3 &28.4    \\
$\phi_{\mathcal{C}_{J}^{0},\mathcal{Z}_{J}^{0}}^{\prime\prime}$
&Proposed  AMA
& $\left\{ \mathcal{C}_{J}^{1}, \mathcal{Z}_{J}^{1} \right\}$
&30.4   &29.9 &30.1   \\
\bottomrule
\end{tabular}
\end{table}

\section{Conclusions}
\label{sec:con}
In this paper we proposed a novel approach to look at the steganographic problem; namely, we proposed to embed the stego message while simultaneously taking into account the necessity of countering an advanced CNN-based steganalyzer.
Such an aim is achieved by introducing a new adversarial embedding method,
which takes both data embedding and adversarial operation into account.
A practical steganographic scheme, AMA, which generates adversarial stego images
with minimum alteration,
has been illustrated  to counter a deep learning based targeted steganalyzer.
The extensive experiments we have carried out permitted us to reach the following conclusions:
\begin{enumerate}
  \item When the targeted steganalyzer is accessible by the steganographer but the steganalyst is unaware of the adversary operation, a high missed detection rate can be achieved by AMA to counter the targeted steganalyzer.
  \item When the steganalyst is aware of the adversarial embedding, and uses adversarial stego images to re-train the steganalyzer, the proposed AMA leads to a higher detection error rate compared to the
        state-of-the-art baseline steganographic scheme, for both targeted and non-targeted steganalyzers.
  \item When both the steganographer and the steganalyst iteratively
        adjust their strategies according to the updated knowledge about the other side,
        the one who makes the last move has a great advantage.
\end{enumerate}

Our approach to adversarial embedding shows a promising way to enhance steganographic security, still there are several unsolved issue to consider.
To start with, the proposed AMA scheme uses only the signs of the gradients.
It worths investigating whether the amplitudes of the gradients can also be helpful.
Besides, the foundation of AMA is the accessibility of the gradients backpropagated from the steganalyzer.
It is worth studying how to counter targeted steganalyzers which do not backpropagate gradients to the input, such as those with hand-crafted features.
Furthermore, for a complete characterization of the interplay between the steganographer and the steganalyst,
it would be interesting to resort to a game-theoretic formulation of the problem \cite{Barni2013a,Barni2013,Barni2014}.



\end{document}